\documentclass{elsart}

\usepackage{natbib}

 \usepackage{graphicx}

\usepackage{amssymb}

\begin{document}

\begin{frontmatter}



\title{Modular organization of cancer signaling networks is associated with patient survivability}


\author[KIT,corr]{Kazuhiro Takemoto}
\ead{takemoto@bio.kyutech.ac.jp}
\author[KIT]{Kaori Kihara}

\address[KIT]{Department of Bioscience and Bioinformatics, Kyushu Institute of Technology, Kawazu 680-4, Iizuka, Fukuoka 820-8502, Japan}



\corauth[corr]{Corresponding author.}

\begin{abstract}
Molecular signaling networks are believed to determine cancer robustness.
Although cancer patient survivability was reported to correlate with the heterogeneous connectivity of the signaling networks inspired by theoretical studies on the increase of network robustness due to the heterogeneous connectivity, other theoretical and data analytic studies suggest an alternative explanation: the impact of modular organization of networks on biological robustness or adaptation to changing environments.
In this study, thus, we evaluate whether the modularity--robustness hypothesis is applicable to cancer using network analysis.
We focus on 14 specific cancer types whose molecular signaling networks are available in databases, and show that modular organization of cancer signaling networks is associated with the patient survival rate.
In particular, the cancers with less modular signaling networks are more curable.
This result is consistent with a prediction from the modularity--robustness hypothesis.
Furthermore, we show that the network modularity is a better descriptor of the patient survival rate than the heterogeneous connectivity.
However, these results do not contradict the importance of the heterogeneous connectivity.
Rather, they provide new and different insights into the relationship between cellular networks and cancer behaviors.
Despite several limitations of data analysis, these findings enhance our understanding of adaptive and evolutionary mechanisms of cancer cells.
\end{abstract}


\end{frontmatter}

\section{Introduction}
\label{sec:intro}
Cancer is a complex and robust system; thus, it may remain an incurable disease despite the efforts to develop effective anticancer therapies \citep{Kitano2004,Tian2011}.
Understanding of the origin of cancer robustness is an important topic of scientific inquiry not only for researchers in the field of basic biology but also for investigators in medical research.

Network biology \citep{Barabasi2004,Albert2005} and network medicine \citep{Barabasi2011,Cho2012} are helpful for untangling complex systems such as cancers.
The biomolecules of living organisms, such as proteins and metabolites, undergo several interactions and chemical reactions, which lead to the occurrence of various life phenomena \citep{Barabasi2004,Albert2005,Alon2006}.
These interactions can be represented in the form of networks or graphs.

Cancer behaviors are governed and coordinated by these interactions between biomolecules (i.e., cancer signaling networks) \citep{Kitano2004,Tian2011,Dreesen2007}.
In recent years, several new technologies and high-throughput methods have generated a massive quantity of data on signaling networks, thus, the understanding of cancer signaling networks is progressively becoming clearer.
In addition, the data on signaling networks are accumulated in several databases such as the Kyoto Encyclopedia of Genes and Genomes (KEGG) \citep{Kanehisa2012}.
As a result, investigators have been able to actively carry out comprehensive data analyses in an ongoing attempt to shed light on the understanding of cancer robustness.

The relationship between network structures and their robustness is well investigated in network biology (or network science \citep{Albert2002,Barabasi2013}, in general).
Especially, it is well known that heterogeneous connectivity (or scale-freeness), which indicates that a few nodes (hubs) integrate numerous nodes while most of the remaining nodes do not, promotes the network robustness against random failures because of hubs, but it leads to the fragility of networks under the condition of targeted attacks to hubs \citep{Albert2000}.
The related works are summarized by \citet{Cohen2010}.

Inspired by these previous studies, \citet{Breitkreutz2012} focused on 13 types of cancers, which are available in the KEGG database, and they found that the patient survivability, which is interpreted as cancer vulnerability because more robust cancers may be more incurable, is correlated with the degree of heterogeneous connectivity of cancer signaling networks.

However, this conclusion has several limitations.
Especially, the prostate cancer was excluded when investigating a correlation between the patient survivability and heterogeneous connectivity of cancer signaling networks because it has minimal vasculature and morphologically distinct.
\citet{Breitkreutz2012} reported that such a negative correlation is not concluded when the prostate cancer also included, suggesting a limitation of heterogeneous connectivity as a descriptor of the patient survivability.

In addition to this, biological robustness is described in a different context: modularity, which, in essence, reflects the deconstruction of a network into dense, and yet, weakly interconnected subnetworks \citep{Fortunato2010}.
Modularity may be related to robustness or capability of quickly adapting to changing environments (i.e., modularity--robustness hypothesis) \citep{Hartwell1999}.
For example, \citet{Kashtan2005} showed that modular networks spontaneously evolve when the evolutionary goal changes over time in a manner that preserves the same subgoals but in different permutations using a theoretical model. 
Similarly, \citet{Lipson2002} suggested that changing environments can promote modularity.
\citet{Hintze2008} showed that modularity evolves in biological networks (metabolic networks in this study) in order to deal with a multitude of functional goals, with a degree depending on environmental variability.
Moreover, \citet{Samal2011} also derived similar conclusions on the relationship between metabolic network modularity and changes in the chemical environment, which they specifically defined as the availability and source of carbon-based molecules, using flux balance analysis.
The conclusion that environmental variability (or changing environments) promotes network modularity is partially supported by the data analysis of real-world biological (metabolic) networks \citep{Parter2007}.
These findings imply that more modular networks are more robust (i.e., have a potential of the adaptation to changing environments).
In this context, a functional and/or evolutionary goal is interpreted as survival of cancer cells under a condition such as drug dosage and radiation exposure.
Taken together, a signaling network of a more incurable cancer is expected to show a higher modularity.

In this study, therefore, we evaluate whether the hypothesis that network modularity is related to robustness is applicable to cancers using network analysis, and show that network modularity is a better descriptor of patient survivability.
Moreover, we discuss insights into cancer evolution (environmental adaptation) and treatments from a viewpoint of modularity.

\section{Methods}
\subsection{Cancer signaling networks and patient survivability}
According to the list of cancer types (i.e., Table 1) in the previous study by \citet{Breitkreutz2012}, we manually downloaded the KGML (KEGG Markup Language) files containing the signaling network data of 14 cancer types on June 25, 2012 from the KEGG database \citep{Kanehisa2012}, and constructed the cancer signaling networks in which nodes and edges are proteins and relations between proteins such as protein--protein interactions and signaling flows.
Although the cancer signaling networks shown in the KEGG database have directed relationships (i.e., edges), the direction of edges is neglected (i.e., the signaling networks are represented as undirected networks) in this study, as in the previous study by \citet{Breitkreutz2012}, because they are not required when calculating the following network parameters.

In addition to this, we obtained the average 5-year survival rate of cancer patients according to the previous study by \citet{Breitkreutz2012}.
The survival rate of cancer patients was originally extracted from the Surveillance Epidemiology and End Results (SEER) Program database \citep{Jemal2013} (http://seer.cancer.gov/), which provides information on cancer statistics compiled by the National Cancer Institute. 

\subsection{Network parameters}
The network parameters, investigated in this study, are explained.
\subsubsection{Degree entropy}
\citet{Breitkreutz2012} focused on the degree entropy for measuring network complexity.
The degree entropy is a kind of {\it graph entropy}, which is often used to evaluate network complexity (reviewed by \citet{Dehmer2011} and \citet{Simonyi1995}).
The graph entropy of network $G$ is based on Shannon's entropy, and it is conceptually defined as $\bar{I}(G)=-\sum_{i=1}^n\frac{|X_i|}{|X|}\log \frac{|X_i|}{|X|}$, where $|X|$ corresponds to a network invariant such as the total number of nodes or the total number of edges.
The network is divided into $n$ subsets, based on a given criterion, and the value $|X_i|$ denotes the cardinality of subset $i$.

The degree entropy $H$ is a simple example of graph entropy, and it is based on the node degree \citep{Rashevsky1995}.
Let $N_k$ be the number of nodes with degree $k$; the degree entropy $H$ is given as
\begin{equation}
H=-\sum_{k=0}^{N-1}\frac{N_k}{N}\log \frac{N_k}{N},
\end{equation}
where $N$ is the total number of nodes.
Since $N_k/N=P(k)$ (i.e., the degree distribution), this equation is rewritten as $H=-\sum_{k=0}^{N-1}P(k)\log P(k)$.
That is, the degree entropy $H$ characterizes the degree of heterogeneity in a network.

\subsubsection{Network modularity}
The modularity of networks is often measured using the $Q$-value, which is widely used for investigating network modularity (e.g., reviewed by \citet{Fortunato2010}).
The network modularity $Q$ is defined as the fraction of edges that lie within, rather than between, modules relative to that expected by chance as follows:
\begin{equation}
Q=\frac{1}{2E}\sum_{ij}\left[A_{ij}-\frac{k_ik_j}{2E}\right]\delta(c_i,c_j),
\end{equation}
where $\delta(c_i,c_j)=1$ if nodes $i$ and $j$ belong to the same module and $0$ otherwise.
$E$ corresponds to the number of edges, and $A_{ij}$ is an adjacency matrix.
$k_i$ denotes the number of neighbors (i.e., node degree) of node $i$.

A network with a higher $Q$ indicates a higher modular structure.
Thus, we have to find the global maximum $Q$ over all possible divisions.
Since it is hard to find the optimal division with the maximum $Q$ in general, approximate optimization techniques are required \citep{Fortunato2010}.  
In this study, an algorithm based on simulated annealing \citep{Reichardt2006} was used for finding the maximum $Q$ in order to avoid the resolution limit problem in community (or module) detection \citep{Fortunato2007,Fortunato2010} as much as possible.
The maximum $Q$ is defined as the network modularity of cancer signaling networks. 

To allow the comparison of network modularity with networks of different size and connectivity, we used the normalized network modularity value $Q_m$ based on the previous studies by \citep{Parter2007,Takemoto2012,Takemoto2011,Takemoto2013}, which was defined as:
\begin{equation}
Q_m=\frac{Q_{\mathrm{real}}-Q_{\mathrm{rand}}}{Q_{\max}-Q_{\mathrm{rand}}},
\end{equation}
where $Q_{\mathrm{real}}$ is the network modularity of a real-world signaling network and $Q_{\mathrm{rand}}$ is the average network modularity value obtained from 10000 randomized networks constructed from its real-world network.
$Q_{\max}$ was estimated as: $1-1/M$, where $M$ is the number of modules in the real network.

Randomized networks were generated from a real-world network using the edge-rewiring algorithm \citep{Maslov2002}.
This algorithm generates a random network by rewiring 2 randomly selected edges until the rewiring of all edges is completed.
For example, consider 2 edges, A--B and C--D, where the letters and lines are nodes and edges, respectively. Through this edge-rewiring algorithm, the edges A--D and C--B are obtained (see the study by \citet{Maslov2002} for details).

\subsubsection{Clustering coefficient}
The clustering is also considered because its concept is similar to that of the network modularity.
To measure the clustering effects, the clustering coefficient \citep{Albert2002,Barabasi2004,Watts1998} was proposed.
This measure denotes the density among neighbors of node $i$, and is defined as the ratio of the number of edges among the neighbors to the number of all possible connections among the neighbors: 
\begin{equation}
C_i=\frac{2M_i}{k_i(k_i-1)},
\end{equation}
$M_i$ is the number of edges among neighbors of node $i$.
In this study, we focused on the overall tendency of clustering that is measured by the average clustering coefficient: $C=[1/N]\sum_{i=1}^N C_i$.
Note that $C$ only focused on the clustering effect among neighboring nodes unlike the network modularity $Q$.

In addition, the largest connected component (giant component) was extracted from each cancer signaling network after adding edges between proteins belonging to the same protein complex (i.e., protein--protein interactions) in order to obtain more accurate calculations of $Q$ and $C$ and to avoid bias from small isolated components.
Especially, the module detection algorithm requires connected undirected networks.

\subsection{Statistical test}
For measuring statistical dependence between the 5-year survival rate and network parameters, we used the Spearman's rank correlation coefficient $r_s$, which is a non-parametric measure (i.e., it is relatively robust to outliers and can be also consider nonlinear relationships), and its $P$-value $p$.

\section{Result}
\label{sec:result}
\begin{figure}[tbhp]
\begin{center}
	\includegraphics{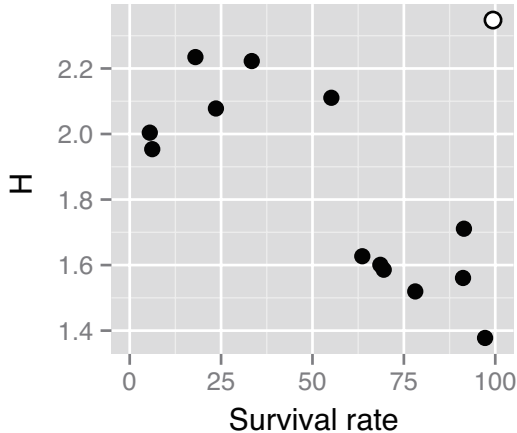}  
	\caption{
	Correlation between the degree entropy $H$ and the 5-year survival rate.
	A significant correlation is not concluded (Spearman's rank correlation coefficient $r_s=-0.41$ and $P$-value $p=0.15$) when considering all 14 cancer types; however, a significant negative correlation is observed when the prostate cancer (open circle) is excluded ($r_s=-0.76$ and $p=0.0036$).
	}
	\label{fig:deg_entropy}
\end{center}
\end{figure}

A correlation between the degree entropy $H$ and the 5-year survival rate of patients was re-confirmed (Fig. \ref{fig:deg_entropy}).
As mentioned by \citet{Breitkreutz2012}, the degree entropy shows a negative correlation with the survival rate when the prostate cancer is neglected.
This result suggests that the signaling networks of cancers with higher patient survivability show a lower heterogeneous connectivity.
In general, since cancer treatments are interpreted as targeted attacks to hub proteins such as p53 \citep{Lane2010}, a famous cancer-related hub protein, such target proteins may be easily found in networks with a higher heterogeneous connectivity.
On the other hand, target proteins may be hardly identified in networks with a lower heterogeneous connectivity because such networks have striking structural properties like random networks.

However, this discussion is debatable because a negative correlation between the degree entropy $H$ and the survival rate is not concluded when considering all 14 cancer types (Fig. \ref{fig:deg_entropy}) although $r_s$ is relatively robust to outliers.
Especially, the prostate cancer has a higher patient survivability than expected from the degree entropy $H$.

\begin{figure}[tbhp]
\begin{center}
	\includegraphics{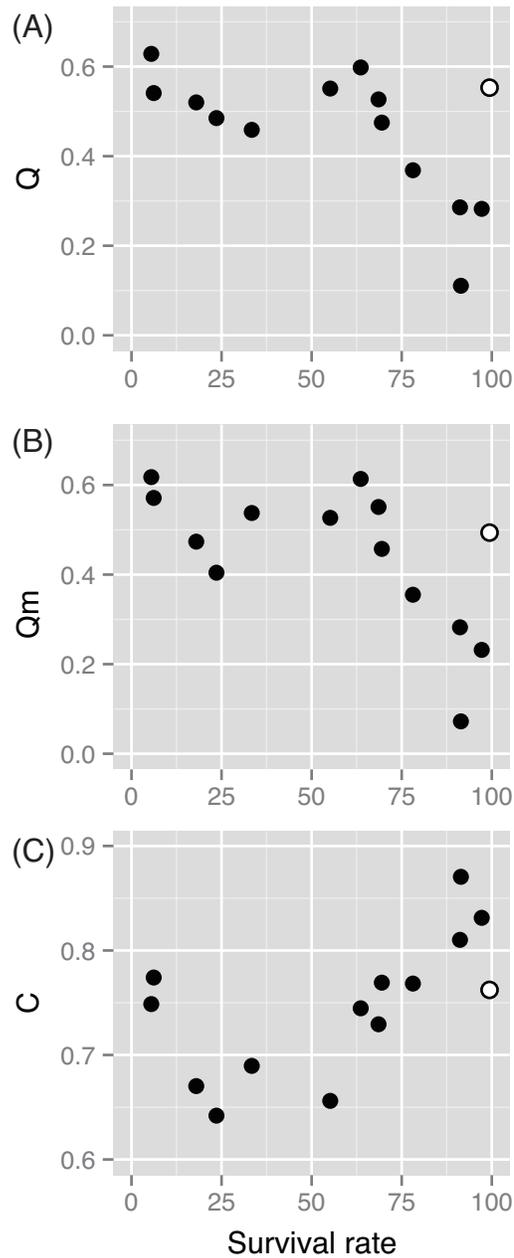}  
	\caption{
	Correlations of the 5-year survival rate with the modularity $Q$ (A) (Spearman's rank correlation coefficient $r_s=-0.50$ and $P$-value $p=0.069$), normalized modularity $Q_m$ (B) ($r_s=-0.653$ and $p=0.014$), and the average clustering coefficient $C$ (C) ($r_s=0.57$ and $p=0.036$).
	When the prostate cancer (open circle) is excluded, the 5-year survival rate shows a more significant correlation with $Q$ ($r_s=-0.77$ and $p=0.0033$), $Q_m$ ($r_s=-0.76$ and $p=0.0036$), and $C$ ($r_s=0.61$ and $p=0.030$), respectively.  
	}
	\label{fig:modularity}
\end{center}
\end{figure}

On the other hand, the modularity-related network parameters are more robustly correlated with the 5-year survival rate of patients (Fig. \ref{fig:modularity}).
In particular, the normalized network modularity $Q_m$ shows a significant negative correlation with the survival rate (Fig. \ref{fig:modularity}B) even if the prostate cancer is also considered although the prostate cancer still seems to be an outlier.
Note that the modularity value was normalized to allow the comparison of the network modularity with networks of different sizes and connectivity, which strongly affect $Q$.
Thus, there was no correlation of $Q_m$ with the number of nodes ($N$) (Spearman's rank correlation coefficient $r_s = 0.27$, $P$-value $p = 0.35$) or the number of edges ($E$) ($r_s = -0.015$, $p = 0.96$). Furthermore, $Q_m$ was not correlated with average degree (i.e., $2E/N$) ($r_s = -0.077$, $p = 0.80$).
The normalization is an important procedure because it is slightly difficult to conclude a negative correlation between the original modularity $Q$ and the survival rate (Fig. \ref{fig:modularity}A).

The average clustering coefficient $C$ shows a positive correlation with the patient survivability (Fig. \ref{fig:modularity}C).
This tendency is not conflict with the negative correlation between $Q_m$ and the survival rate because $C$ only characterizes the clustering effect among neighbors.
For example, a complete graph shows the high $C$ (i.e., $C=1$); however, the modularity $Q$ is 0; thus, an opposite tendency may be observed between $Q$ and $C$. 
In this manner, these network parameters indicate modular structure of networks in a different context.
In fact, the relationship between the normalized modularity $Q_m$ and the clustering coefficient $C$ was not concluded ($r_s=-0.46$ and $p=0.10$).

\begin{figure}[tbhp]
\begin{center}
	\includegraphics{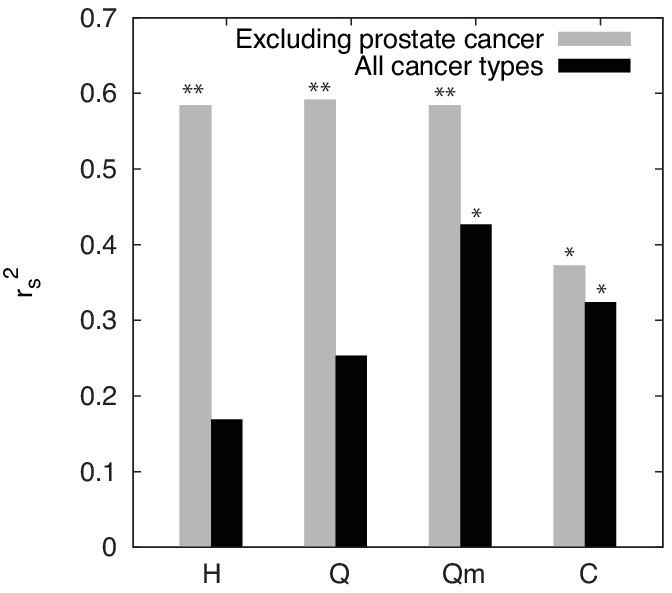}  
	\caption{
	Squared Spearman's rank correlation coefficient $r_s^2$ of the 5-year survival rate with the degree entropy $H$, the modularity $Q$, the normalized modularity $Q_m$, and the average clustering coefficient $C$.
	One asterisk and two asterisks on the top of bars indicate that $P$-value for $r_s$ is less than 0.05 and less than 0.01, respectively.
	The black bars and gray bars correspond to $r_s^2$ for all cancer types and $r_s^2$ observed when the prostate cancer is excluded, respectively.
	}
	\label{fig:bar_rs2}
\end{center}
\end{figure}

The correlations of $Q$, $Q_m$, and $C$ with the 5-year survival rate are more significant if the prostate cancer is neglected (Figs. \ref{fig:modularity} and \ref{fig:bar_rs2}).
Especially, the squared correlation coefficient $r_s^2$ is almost similar between the degree entropy $H$ and the network modularity values $Q$ or $Q_m$.
This result suggests that the network modularity ($Q_m$, in particular) is a better descriptor for explaining the cancer patient survivability.  

\begin{figure}[tbhp]
\begin{center}
	\includegraphics{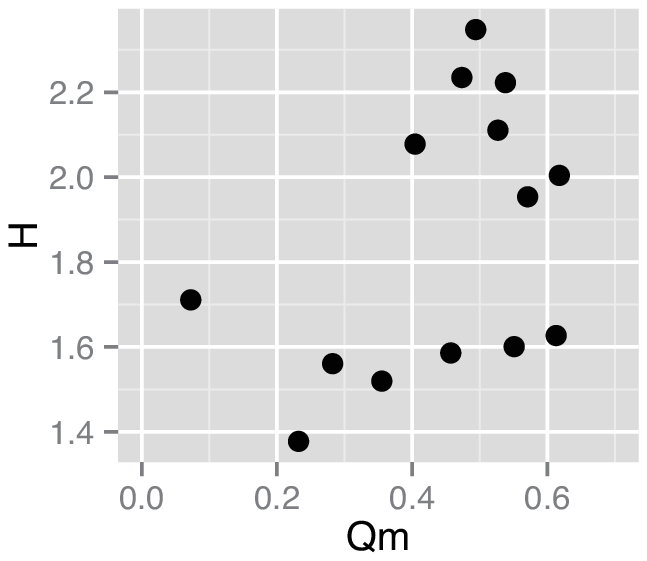}  
	\caption{
	No correlation between the degree entropy $H$ and the normalized modularity $Q_m$ (Spearman's rank correlation coefficient $r_s=0.39$ and $P$-value $p=0.17$).
	}
	\label{fig:cor_Qm_H}
\end{center}
\end{figure}

The normalized network modularity $Q_m$ and the degree entropy $H$ are different structural properties because of no correlation between them (Fig. \ref{fig:cor_Qm_H}).
On the other hand, the clustering coefficient $C$ is related to the degree entropy $H$ (Spearman's rank correlation coefficient $r_s=-0.61$ and $P$-value $p=0.022$) because these network parameters are computed based on node degrees.
This result implies that the positive correlation between $C$ and the survival rate likely to be a side effect of the negative correlation between $H$ and the survival rate.
Thus, we concluded that the modularity value $Q_m$ is the best descriptor in this study.

To evaluate the contribution of each network property to the patient survival rates, we conducted a kernel partial least squares (PLS) regression analysis \citep{Rosipal2001} using a statistical software R version 3.0.0 \citep{r-team} and with its function {\tt plsr}, which is available in the R package {\tt pls} version 2.3-0.
The PLS are more powerful than than other multivariate analysis techniques.
In particular, the PLS are powerful for extracting relative factors for objective variables from a large number of explanatory variables when a high level of multicollinearity is observed among the explanatory variables.

\begin{figure}[tbhp]
\begin{center}
	\includegraphics{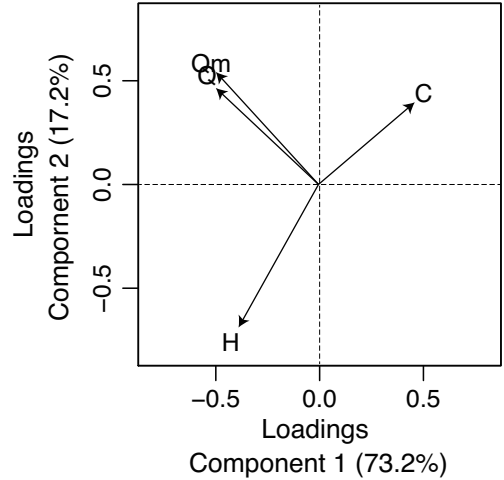}  
	\caption{Loadings plot in the partial least squares regression analysis.
	The percentages in parentheses correspond to the proportions of variance explained in the components 1 and 2, respectively.}
	\label{fig:biplot}
\end{center}
\end{figure}

The loadings plot (Fig. \ref{fig:biplot}) suggests the patient survival rates is mainly explained by the component 1 in which the loadings are $H=-0.43$, $Q=-0.54$, $Q_m=-0.52$, and $C=0.50$, respectively.
This result indicates that the modularity-related measures are more dominant than the degree entropy $H$ for explaining the patient survival rates, and it is a more convincing evidence of the conclusion derived from the correlation analysis (Fig. \ref{fig:bar_rs2}).
However, we conclude that both network properties can contribute to the patient survivability because the difference of the loadings between $H$ and the modularity-related measures is not so large.
Rather, we emphasize the loading vectors are different between the modularity-related measures and $H$.
This result represents the different tendency of the contribution to the cancer patient survivability between the network parameters, and it implies that modularity-related measures are essentially different from $H$.
This conclusion is also supported by no correlation between $Q_m$ and $H$ (Fig. \ref{fig:cor_Qm_H}).

\section{Discussions}
In summary, we showed that the cancer patient survivability is correlated with not only the degree of heterogeneous connectivity (Fig. \ref{fig:deg_entropy}) but also the network modularity (Fig. \ref{fig:modularity}), as expected from the hypothesis that modularity enhances robustness (e.g., capability of quickly adapting to changing environments), suggested by several theoretical and data analytic studies.
Especially, the network modularity more robustly explains the cancer patient survivability than the heterogeneous connectivity does (Fig. \ref{fig:bar_rs2}) because a correlation between the network modularity $Q_m$ and the survival rate is still concluded even if the prostate cancer is considered.
In the previous study by \citet{Breitkreutz2012}, the prostate cancer was neglected because it is an exception due to their morphological differences.
Thus, the network modularity is more widely useful for understanding cancer behaviors than the heterogeneous connectivity from a viewpoint of molecular signaling networks.
However, this result does not contradict the importance of the heterogeneous connectivity in the prediction of cancer survival rates.
The PLS analysis (Fig. \ref{fig:biplot}) suggests the contribution of both heterogeneous connectivity and modularity to the patient survivability although the degree of the contribution of the modularity is higher than that of the heterogeneous connectivity.
Rather, the PLS analysis and the correlation analysis (Fig. \ref{fig:cor_Qm_H}) suggest that the network modularity plays a different role from the heterogeneous connectivity on the patient survivability.

Thus, this finding provides new and different insights into cancer robustness from the heterogeneous connectivity.
In particular, modularity may facilitate an adaptation to changing environments \citep{Hartwell1999}.
Discrete modules in systems (e.g., networks) may archive particular functions; thus, systems are expected to acquire more modules when they have to robustly respond (e.g., cancer cells grow and survive) under more various conditions.
Thus, a cancer with more modular signaling networks is more robust to multiple treatments such as the dosage of multiple drugs and radiation exposure.
This interpretation is consistent with multidrug resistance in cancer \citep{Gillet2010}.
Theoretical studies on the relationship between modularity and adaptation to changing environments, as explained in Sec. \ref{sec:intro}, may be applicable to a deeper understanding of cancer behaviors and treatments.

The network modularity improves the current understanding of cancer robustness obtained from the heterogeneous connectivity.
Although the heterogeneous connectivity suggests the existence of hub proteins such as p53 and its relationship with network robustness \citep{Albert2000,Cohen2010}, recent studies encourage a reconsideration of the importance of hub proteins.
For example, \citet{Han2004} showed that hub proteins can be classified into 2 types despite a criticism to this dichotomy \citep{Agarwal2010}: party hubs that coordinate a specific functional component and date hubs that play a role of intermediates between different specific functional modules.
In addition to this, they found that the effect of hub removals on cellular networks is different between party hubs and date hubs.
In particular, the removal of date hubs leads to a more immediate collapse of cellular networks than that of party hubs.
This findings implies the importance of hubs bridging between different network modules.
Such an importance is also suggested by \citet{Guimera2005} and \citet{Yu2007}. 
Therefore, a functional cartography method \citep{Guimera2005}, revealing a patterns of intra- and inter-module connections in complex networks, and the concept of bottlenecks \citep{Yu2007}, key connectors with functional properties, are more useful for finding target proteins in cancer therapies.

Network modularity is also related to gene duplication.
Using generative models for complex networks, \citet{Hallinan2004} and \citet{Ward2007} showed that networks can acquire modular organization through gene duplication events.
The similar conclusion is derived from the duplication-divergence model \citep{Vazquez2003} and the Dorogovtsev-Mendes-Samukhin model \citep{Dorogovtsev2001}.
This result is important for a deeper understanding of the relationship between the modularity and the adaptation to changing environments.
In particular, the extent of gene duplication is correlated with habitat variability \citep{Makino2012}, which is related to the capability of quickly adapting to changing environments.
Gene duplications are believed to increase mutational robustness because they lead to functional redundancy in biological systems \citep{Wagner2008}.
In fact, several studies suggests that biological components (e.g., proteins) belonging to the same module in cellular networks have similar or related functions (e.g., reviewed by \citet{Cho2012}).
That is, gene duplications are microscopic mechanisms for biological robustness.
They may lead to modularity at a higher level of organization (i.e., cellular network level), and may finally establish an adaptation to changing environments at a phenotypic level.
This speculation suggests the impact of gene duplications on cancer robustness; thus, such an impact may remain to be investigated using available databases such as the Duplicated Genes Database \citep{Ouedraogo2012}. 

Gene duplications facilitate the heterogeneous connectivity (or scale-freeness) in gene regulatory networks \citep{Teichmann2004} and protein networks \citep{Pastor-Satorras2003} because they result preferential attachments (`rich-gets-richer' mechanisms) because nodes with many neighbors tend to obtain more neighbors when considering such mechanisms.
Thus, gene duplications lead to both the heterogeneous connectivity and network modularity.
According to these theories, however, functional divergences resulting edge additions, removals, rewiring, also influence the heterogeneous connectivity.
Thus, it is not necessarily that the heterogeneous connectivity is associated with the network modularity.
In fact, such a relationship was not observed (Fig. \ref{fig:cor_Qm_H})

Although many studies support the modularity--robustness hypothesis, there are some criticisms to this hypothesis.
For example, \citet{Sole2008} qualitatively showed that cellular networks can spontaneously acquire modular organization through evolutionary events such as duplication and divergence using a growing network model.
\citet{Clune2013} demonstrated that the evolution of modularity does not depend on changing environments, but it is related to the selection pressure to reduce the cost of connections between network nodes. 
Furthermore, \citet{Takemoto2012} represented, quantitatively, that network modularity can arise from simple growth processes, without consideration of an adaptation to environmental changes.
\citet{Holme2011} revealed that the network modularity is not a general principle for either strengthening or weakening robustness using a mass-action kinetic model.
Some data analytic studies also support limited effect of network modularity on capability of quickly adapting to changing environments. 
For example, \citet{Takemoto2011} found that growth conditions, trophic requirement, and optimal growth temperature affect network modularity rather than environmental variability.
\citet{Zhou2012} also derived the similar conclusion using a larger dataset.
Furthermore, \citet{Takemoto2013} pointed out that the previously observed increase in network modularity due to habitat variability was probably due to a lack of available data on cellular networks.
Although the limited effect of network modularity is still debatable because it was concluded in biological networks other than molecular signaling networks (metabolic networks in most cases), it may be a fact that the explanation of cancer patient survivability using the modularity--robustness hypothesis has limitations.
When a greater variety of cancer signaling networks will be available in the future, we may observe exceptions in which the patient survivability cannot be explained using molecular signaling networks in the context of network modularity.

The definition of network modularity is also controvertible.
The conclusion in this study is limited in the context of network modularity, which is only identified based on network topology.
In particular, it is pointed out that the definition of modularity might not be topologically intuitive because of the locality and limited resolution \citep{Fortunato2007} although we avoided such limitations as much as possible using a module detection algorithm based on simulated annealing.
Alternative quantitative functions for community partition (e.g., the modularity density $D$ \citep{Li2008}) and methods based on link communities (e.g., \citep{Ahn2010}), in which a $D$-based quantitative function, are used may be useful to avoid these limitations because they show a better prediction of biologically functional modules or categories.
In this study, however, we did not consider these approaches because the quantitative functions (i.e., $D$) is not suitable for comparing network modularity with networks of different size and connectivity (i.e., the normalization method is not established).
In addition to this, our analysis has more general limitations, as do many other works on network analyses: limited knowledge of biomolecular interactions (i.e., missing links) and direction of cellular interaction such as signaling flows.

Although data analysis has several limitations, these findings provide new insights into the relationship between cellular networks (a microscopic view) and phenotypes (a macroscopic view) in cancer, and they enhance our understanding of adaptive and evolutionary mechanisms of cancer cells.
We believe that these findings are also helpful for network-based cancer treatments. 

\section*{Acknowledgments}
This work was supported by a Grant-in-Aid for Young Scientists (A) from the Japan Society for the Promotion of Science (no. 25700030).
K.T. was partly supported by Chinese Academy of Sciences Fellowships for Young International Scientists (no. 2012Y1SB0014), and the International Young Scientists Program of the National Natural Science Foundation of China (no. 11250110508).


\begin{thebibliography}{00}

\bibitem[Ahn et al., 2010]{Ahn2010}
Ahn, Y.-Y., Bagrow, J.P., Lehmann, S., 2010. Link communities reveal multiscale complexity in networks. Nature 466, 761--764.

\bibitem[Albert, 2005]{Albert2005}
Albert, R., 2005. Scale-free networks in cell biology. J. Cell Sci. 118, 4947--4957.

\bibitem[Albert, 2002]{Albert2002}
Albert, R., Barab\'asi, A.-L., 2002. Statistical mechanics of complex networks. Rev. Mod. Phys. 74, 47--97.

\bibitem[Albert et al., 2000]{Albert2000}
Albert, R., Jeong, H., Barab\'asi, A.-L., 2000. Error and attack tolerance of complex networks. Nature 406, 378--482.

\bibitem[Alon, 2006]{Alon2006}
Alon, U., 2006. An Introduction to Systems Biology: Design Principles of Biological Circuits, Chapman \& Hall/CRC, Florida.

\bibitem[Agarwal et al., 2010]{Agarwal2010}
Agarwal, S., Deane, C.M., Porter, M.A., Jones, N.S., 2010. Revisiting date and party hubs: novel approaches to role assignment in protein interaction networks. PLoS Compt. Biol. 6, e1000817.

\bibitem[Barab\'asi, 2013]{Barabasi2013}
Barab\'asi, A.-L., 2013. Network science. Philos. Trans. R. Soc. London, Ser. A 371, 20120375.

\bibitem[Barab\'asi and Oltvai, 2004]{Barabasi2004}
Barab\'asi, A.-L. \& Oltvai, Z.N., 2004. Network biology: understanding the cell's functional organization. Nat. Rev. Genet. 5, 101--113.

\bibitem[Barab\'asi et al., 2011]{Barabasi2011}
Barab\'asi, A.-L., Gulbahce, N., Loscalzo, J., 2011. Network medicine: a network-based approach to human disease. Nat. Rev. Genet. 12, 56--68.

\bibitem[Breitkreutz et al., 2012]{Breitkreutz2012}
Breitkreutz, D., Hlatky, L., Rietman, E., Tuszynski, J.A., 2012, Molecular signaling network complexity is correlated with cancer patient survivability. Proc. Natl. Acad. Sci. USA 109, 9209--9212.

\bibitem[Cho et al., 2012]{Cho2012}
Cho, D.-Y., Kim, Y.-A., Przytycka, T.M., 2012. Network biology approach to complex diseases. PLoS Compt. Biol. 8, e1002820.

\bibitem[Clune et al., 2013]{Clune2013}
Clune, J., Mouret, J.-B., Lipson, H., 2013. The evolutionary origins of modularity. Proc. R. Soc. B. 280, 20122863.

\bibitem[Cohen and Havlin, 2010]{Cohen2010}
Cohen, R., Havlin, S., 2010. Complex Networks: Structure, Robustness and Function. Cambridge University Press, Cambridge.

\bibitem[Dehmer and Mowshowitz, 2011]{Dehmer2011}
Dehmer, M, Mowshowitz, A., 2011. A history of graph entropy measures. Inform. Sci. 181, 57--78.

\bibitem[Dorogovtsev et al., 2001]{Dorogovtsev2001}
Dorogovtsev, S.N., Mendes, J.F.F., Samukhin, A.N., 2001. Size-dependent degree distribution of a scale-free growing network. Phys. Rev. E 63, 062101.

\bibitem[Dreesen and Brivanlou, 2007]{Dreesen2007}
Dreesen, O., Brivanlou, A.H., 2007. Signaling pathways in cancer and embryonic stem cells. Stem Cell Rev. 3, 7--17.

\bibitem[Fortunato, 2010]{Fortunato2010}
Fortunato, S., 2010. Community detection in graphs. Phys. Rep. 486, 75-174.

\bibitem[Fortunato and Barth\'elemy, 2007]{Fortunato2007}
Fortunato, S., Barth\'elemy, M., 2007. Resolution limit in community detection. Proc. Natl. Acad. Sci. USA 104, 39--41.

\bibitem[Gillet and Gottesman, 2010]{Gillet2010}
Gillet, J.P., Gottesman, M.M., 2010. Mechanisms of multidrug resistance in cancer. Methods Mol Biol. 596, 47--76.

\bibitem[Guimer\'a and Amaral, 2005]{Guimera2005}
Guimer\'a, R., Amaral, L.A.N., 2005. Functional cartography of complex metabolic networks. Nature 433, 895--900.

\bibitem[Hallinan, 2004]{Hallinan2004}
Hallinan, J., 2004. Gene duplication and hierarchical modularity in intracellular interaction networks. Biosystems 74, 51--62. 

\bibitem[Han et al., 2004]{Han2004}
Han, J.-D.J., et al., 2004. Evidence for dynamically organized modularity in the yeast protein--protein interaction network. Nature 430, 88--93.

\bibitem[Hartwell et al., 1999]{Hartwell1999}
Hartwell, L.H., Hopfield, J.J., Leibler, S., Murray, A.W., 1999. From molecular to modular cell biology. Nature 402, C47--C52.

\bibitem[Hintze and Adami, 2008]{Hintze2008}
Hintze, A., Adami, C., 2008. Evolution of complex modular biological networks. PLoS Comput. Biol. 4, e23.

\bibitem[Holme, 2011]{Holme2011}
Holme, P., 2011. Metabolic robustness and network modularity: a model study. PLoS ONE 6, e16605.

\bibitem[Jemal et al., 2013]{Jemal2013}
Jemal, A., et al., 2013. Annual report to the Nation on the Status of Cancer, 1975–2009, featuring the burden and trends in Human Papillomavirus (HPV)–Associated cancers and HPV vaccination coverage levels. J. Natl. Cancer Inst. 105, 175--201.

\bibitem[Kanehisa et al., 2012]{Kanehisa2012}
Kanehisa, M., Goto, S., Sato, Y., Furumichi, M., Tanabe, M., 2012. KEGG for integration and interpretation of large-scale molecular data sets. Nucleic Acids Res. 40, D109--D114.

\bibitem[Kashtan and Alon, 2005]{Kashtan2005}
Kashtan, N., Alon, U., Spontaneous evolution of modularity and network motifs. Proc. Natl. Acad. Sci. USA 102, 13773--13778.

\bibitem[Kitano, 2004]{Kitano2004}
Kitano, H., 2004. Cancer as a robust system: implications for anticancer therapy. Nat. Rev. Cancer 4, 227--235.

\bibitem[Lane et al., 2010]{Lane2010}
Lane, D.P., Cheok, C.F., Lain, S., 2010. p53-based cancer therapy. Cold Spring Harb. Perspect. Biol. 2, a001222.

\bibitem[Li et al., 2008]{Li2008}
Li, Z., Zhang, S., Wang, R.S., Zhang, X.S., Chen, L., 2008. Quantitative function for community detection. Phys. Rev. E 77, 036109.

\bibitem[Lipson et al., 2002]{Lipson2002}
Lipson, H., Pollack, J., Suh, N.
On the origin of modular variation. Evolution 56, 1549--1556.

\bibitem[Makino and Kawata, 2012]{Makino2012}
Makino, T., Kawata, M., 2012. Habitat variability correlates with duplicate content of {\it Drosophila} genomes. Mol. Biol. Evol. 29, 3169--3179.

\bibitem[Maslov and Sneppen, 2002]{Maslov2002}
Maslov, S., Sneppen, K., 2002. Specificity and stability in topology of protein networks. Science 296, 910--913.

\bibitem[Ouedraogo et al., 2012]{Ouedraogo2012}
Ouedraogo, M., et al., 2012. The Duplicated Genes Database: identification and functional annotation of co-localised duplicated genes across genomes. PLoS ONE 7, e50653. 

\bibitem[Parter et al., 2007]{Parter2007}
Parter, M., Kashtan, N., Alon, U., 2007.
Environmental variability and modularity of bacterial metabolic networks.
BMC Evol. Biol. 7, 169.

\bibitem[Pastor-Satorras et al., 2003]{Pastor-Satorras2003}
Pastor-Satorras, R., Smith, E., Sol\'e, R.V., 2003. Evolving protein interaction networks through gene duplication. J. Theor. Biol. 222, 199--210.

\bibitem[Rashevsky, 1995]{Rashevsky1995}
Rashevsky, N., 1995. Life information theory and topology. Bull. Math. Biophys. 17, 299--235.

\bibitem[R Core Team, 2013]{r-team}
R Core Team, 2013. R: A language and environment for statistical computing. R Foundation for Statistical Computing, Vienna, Austria. Available: http://www.R-project.org/

\bibitem[Reichardt and Bornholdt, 2006]{Reichardt2006}
Reichardt, J., Bornholdt, S., 2006. Statistical mechanics of community detection. Phys. Rev. E. 74, 016110.

\bibitem[Rosipal and Trejo, 2001]{Rosipal2001}
Rosipal, R., Trejo, L.J., 2001. Kernel partial least squares regression in reproducing kernel Hilbert space. J. Mach. Learn. Res. 2, 97--123

\bibitem[Samal and Wagner, 2011]{Samal2011}
Samal, A., Wagner, A., Martin, O., 2011. Environmental versatility promotes modularity in genome-scale metabolic networks. BMC Syst. Biol. 5, 135. 

\bibitem[Simonyi, 1995]{Simonyi1995}
Simonyi, G., 1995. Graph entropy: a survey. In {\it Combinatorial Optimization} (Eds.: W. Cook, L. Lov\'asz, P. Seymour), DIMACS Series in Discrete Mathematics and Theoretical Computer Science 20, 399--441.

\bibitem[Sol\'e and Valverde, 2008]{Sole2008}
Sol\'e, R.V., Valverde, S., 2008. Spontaneous emergence of modularity in cellular networks. J. R. Soc. Interface 5, 129--133.

\bibitem[Takemoto, 2012]{Takemoto2012}
Takemoto, K., 2012. Metabolic network modularity arising from simple growth processes. Phys. Rev. E. 86, 036107.

\bibitem[Takemoto, 2013]{Takemoto2013}
Takemoto, K., 2013. Does habitat variability really promote metabolic network modularity? PLoS ONE 8, e61348.

\bibitem[Takemoto and Borjigin, 2011]{Takemoto2011}
Takemoto, K., Borjigin, S., 2011. Metabolic network modularity in archaea depends on growth conditions. PLoS ONE 6, e25874.

\bibitem[Teichmann and Babu, 2004]{Teichmann2004}
Teichmann, S.A., Babu, M.M., 2004. Gene regulatory network growth by duplication. Nat. Genet. 36, 492--496.

\bibitem[Tian et al., 2011]{Tian2011}
Tian, T., Olson, S., Whitacre, J.M., Harding, A., 2011. The origins of cancer robustness and evolvability. Integr. Biol. 3, 17--30.

\bibitem[V\'azquez, 2003]{Vazquez2003}
V\'azquez, A., 2003. Growing network with local rules: preferential attachment, clustering hierarchy, and degree correlations. Phys. Rev. E 67, 056104.

\bibitem[Wagner, 2008]{Wagner2008}
Wagner, A., 2008. Gene duplications, robustness and evolutionary innovations. BioEssays 30, 367--373.

\bibitem[Ward and Thornton, 2007]{Ward2007}
Ward, J., Thornton, J., 2007. Evolutionary models for formation of network motifs and modularity in the {\it Saccharomyces} transcription factor network. PLoS Compt. Biol. 3, e198.

\bibitem[Watts and Strogatz, 1998]{Watts1998}
Watts, D.J., Strogatz, S.H., 1998., Collective dynamics of `small-world' networks, Nature 393, 440--442.

\bibitem[Yu et al., 2007]{Yu2007}
Yu, H., Kim, P.M., Sprecher, E., Trifonov, V., Gerstein, M., The importance of bottlenecks in protein networks: correlation with gene essentiality and expression dynamics. PLoS Compt. Biol. 3, e59.

\bibitem[Zhou and Nakhleh, 2012]{Zhou2012}
Zhou, W., Nakhleh, L., 2012. Convergent evolution of modularity in metabolic networks through different community structures. BMC Evol. Biol. 12, 181.

\end{thebibliography}
\end{document}